# Localization of Shear in Saturated Granular Media: Insights from a Multi-Scaled Granular-Fluid Model

Einat Aharonov[1], Liran Goren[2], David W. Sparks[3], and Renaud Toussaint[4]

[1] Institute of Earth Sciences, The Hebrew University of Jerusalem, Jerusalem, Israel, 91940; PH (972) 544-956-907; FAX (972) 4-6388320; email: einatah@cc.huji.ac.il
[2] ETH Zurich, Switzerland; email: liran.goren@erdw.ethz.ch
[3] Texas A&M University, TX, USA; email: sparks@geo.tamu.edu
[4] Institut de Physique du Globe de Strasbourg (IPGS), UMR 7516, Université de Strasbourg/EOST, CNRS; 5 rue Descartes, F-67084 Strasbourg Cedex, France; email: renaud.toussaint@unistra.fr

**ABSTRACT**

The coupled mechanics of fluid-filled granular media controls the behavior of many natural systems such as saturated soils, fault gouge, and landslides. The grain motion and the fluid pressure influence each other: It is well established that when the fluid pressure rises, the shear resistance of fluid-filled granular systems decreases, and as a result catastrophic events such as soil liquefaction, earthquakes, and accelerating landslides may be triggered. Alternatively, when the pore pressure drops, the shear resistance of these systems increases. Despite the great importance of the coupled mechanics of grains-fluid systems, the basic physics that controls this coupling is far from understood.

We developed a new multi-scaled model based on the discrete element method, coupled with a continuum model of fluid pressure, to explore this dynamical system. The model was shown recently to capture essential feedbacks between porosity changes arising from rearrangement of grains, and local pressure variations due to changing pore configurations. We report here new results from numerical experiments of a continuously shearing layer of circular two-dimensional grains, trapped between two parallel rough boundaries. The experiments use a fixed confining stress on the boundary walls, and a constant velocity applied to one of the boundaries, as if this system was the interior of a sliding geological fault filled with 'fault gouge'. In addition, we control the layer permeability and the drainage boundary conditions. This paper presents modeling results showing that the localization of shear (into a narrow shear band within the shearing layer) is strongly affected by the presence of fluids. While in dry granular layers there is no preferred position for the onset of localization, drained systems tend to localize shear on their boundary. We propose a scaling argument to describe the pressure deviations in a shear band, and use that to predict the allowable positions of shear localizations as a function of the fault and gouge properties.

## INTRODUCTION

There is still no consistent, general physical or mathematical formalism that successfully predicts the collective behavior of a large number of discrete grains, even for simple shear of 2D dry granular media, let alone for a case that includes fluid between the grains. The question of a shearing grain layer is very basic; the analog of Couette flow in fluids, and it is relevant for both industrial and earth sciences applications. Here we are mainly interested in a situation relevant to earth sciences, although the setup and questions we pose are quite general. We study the behavior of a fragmented rock layer, termed 'gouge', within a geological fault, which is basically a shearing granular layer. Such granular layers are often the site of earthquake nucleation and rupture, or the layer on which landslide shear motion occurs. The setup discussed here is a granular layer confined between walls, pressed by a confining stress and sheared at constant velocity (see Figure 1).

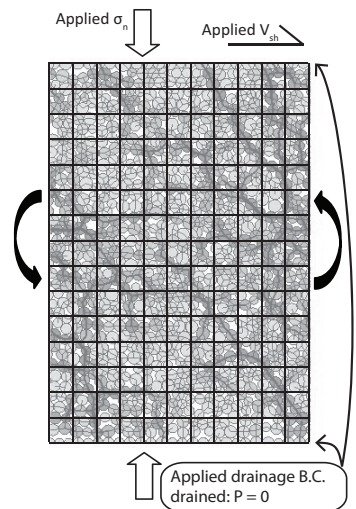

**Figure 1. Grains–fluid simulations setup. In each simulation a collection of grains is packed within a rectangular box with wrap-around boundary conditions along the horizontal direction. Normal stress, $\sigma_n$, and shear velocity, $V_{sh}$, are applied and maintained constant. The spatial and temporal evolution of porosity, $\Phi$, and of pore pressure, $P$, is measured.**

Many previous studies have investigated this setup, and discussed the distribution of shear within the layer. Shear localization into a narrow shear band (about 7-10 grains wide [Oda and Kazma, 1998; Howell et al., 1999]) will occur in continued shearing of dilatant granular materials under uniform stresses and strains [Rudnicki and Rice, 1975]. Mead [1925] calculated that frictional work includes a term due to layer dilation, and suggested that localization occurs when it helps to minimize the frictional work by reducing the dilation component that is expected to be more prominent when the deformation is distributed. This connection between frictional weakening, localization, and dilation was demonstrated in the experiments of Marone [1991] and Rathbun and Marone [2010]. It was also theoretically and numerically demonstrated in Makedonska et al. [2011], using Discrete Element Method (DEM) simulations. It is thus clear that understanding localization is tightly linked to understanding friction of granular layers.

The question of localization is even more complicated when the grain layer is fluid-saturated, as is usually the case for geological faults. The presence of fluid affects slip via the mechanical coupling between granular deformation and pore fluid pressure. The most important aspect of this coupling is the fact that the shear resistance, $\tau$, of a saturated granular material decreases with increasing fluid pressure, $P$, as $\tau \sim \sigma_n - P$, where $\sigma_n$ is the confining stress [Terzaghi, 1943; Boitnott and Scholz, 1990]. This law, called the 'effective stress law', may be used to predict that dilation will stabilize sliding via pore pressure reduction [Segall and Rice, 1995; Moore and Iverson, 2002], while shear induced compaction of under consolidated gouge may lead to extreme weakening and unstable sliding [Blanpied et al., 1992].

This paper uses a new numerical model to study the pattern of localization in fluid-saturated granular layers and compares with localization in dry granular layers. We propose and demonstrate that pore pressure effects are crucial with regards to localization, since pore pressure variations within the layer may lead to a favored locality for localization where shear resistance is minimized.

**MODELING METHOD**

We use a new two-phase two-scale model to simulate the coupled dynamics of grains and pore fluid and specifically to compare dry and wet localization. The model is described in detail in Goren et al.[2010] and [2011], and is summarized only briefly here. This model builds upon the popular DEM, which models a grain layer as a collection of grains (in 2D as disks) interacting with each other via inelastic collisions and frictional forces. The fluid phase is added on a coarser scale (a few grain diameters), with the fluid pressure deviation from hydrostatic value, P, solved on a superimposed Eulerian grid. The fluid does not see the detailed pore geometry between the grains, but instead interacts with a locally averaged porosity and permeability. The solid grains feel pressure gradients interpolated from the fluid grid and added as additional forces. This model was validated in Goren et al. [2011] who presented simulations of several different problems and compared with theory: the effective stress law, sedimentation, fluidization, and most importantly liquefaction onset and conditions (under drained and undrained boundary conditions).

Grains with a range of diameters (from 0.8 mm to 1.2 mm) are randomly packed between two rigid walls made of glued grains. One wall is held fixed. Although no gravity is applied to the system, for convenience we will refer to this wall as the bottom wall, and to the shearing direction as horizontal. The two side boundaries are wrap-around, as in a cylindrical shearing system. In each of the simulations a fixed normal stress, $\sigma_n$, and a fixed shear velocity, $V_{sh}$, are applied (see Table 1).

The current set of simulations with fluid also specify drained fluid boundary conditions at the walls, i.e. $P=0$ outside the granular layer, allowing flow through permeable walls. This mimics a situation where the fault consists of a high permeability breccia that borders a low permeability slipping region in the fault core, which is made of very fine-grained clay-like material. For example, the Nojima Fault core has lower permeability by several orders of magnitude relative to the brecciated zone [Lockner et al., 2009].

**LOCALIZATION IN DRY GRANULAR LAYERS**

Aharonov and Sparks [2002] studied the patterns of localization that occur under various normal stresses and shear velocities in dry granular media, using the DEM without fluid. They found two modes for localization under constant shear velocity, which they termed Fluid-like and Solid-like. The term Fluid-like that they used refers not to pore fluid, but to the transient nature of grain contacts and stress transmission. This mode of localization occurs under very low applied normal stress and under very fast shear velocity. In this case localization occurs at the top boundary, where shear is applied, in a layer consisting of agitated grains that behave as 'jumping beans', and exert high enough dynamic stresses so as to lift the upper wall. This regime is not expected under the high confining stresses occurring in geological faults. Instead a Solid-like mode occurs under higher confining stress, where stresses are supported by stress-chains and not by grain collisions. All the experiments in this paper were conducted at conditions that would fall in the Solid-like mode.

For dry shearing grains, in a Solid-like mode and with no grain-breakage, shear in the system is always localized, but the position of the localized shear bands jumps randomly throughout the layer with time. Shear bands form at a random depth in a layer, persist for a short duration, and then move to another location. Figure 2 shows the behavior of run D10 Dry (see Table 1 for simulation parameters). Because of the wrap-around conditions, deformation in this system localizes into horizontal shear bands that span the system. Therefore it is useful to average grain-properties such as velocity within horizontal layers. The top and middle panels of Figure 2 show the evolution of horizontally-averaged properties: the system is divided into horizontal layers approximately two grain diameters high; each vertical column in the plot shows all horizontal layers at a snapshot in time, and each horizontal row shows the temporal evolution of a particular horizontal layer.

**Table 1. Simulations Parameters**

| Simulation[a] | $\sigma_n$ [MPa] | $V_{sh}$ [m/s] | Internal Permeability[b] [m$^2$] |
|---|---|---|---|
| D6 Wet | 24 | 7.6 | $10^{-12}$ |
| D10 Dry | 24 | 7.6 | - |
| D10 Wet | 24 | 7.6 | $10^{-14}$ |
| D11 Wet | 2.4 | 0.76 | $10^{-14}$ |

[a]All the simulations are about 68.5 grains high and 24 grains wide. We use the viscosity of water, $\eta=10^{-3}$ Pa s, and the compressibility of water, $\beta=4.5\times10^{-10}$ Pa$^{-1}$.
[b]Order of magnitude.

The top panel of Figure 2 shows shear strain rate (variation in average horizontal grain velocity, $V_x$, with depth) $dV_x/dy$. The horizontal axis is non-dimensional time. Dark shadings are high sympathetic shear strain rate (in the direction of applied shear), gray shading shows no shear and light shading shows antithetic shear. The black regions are the location of intense shear, or shear bands. The figure shows that shear bands localize in various places throughout the upper half of the layer, spending only brief periods at each locality. The middle panel shows volume strain rate (variation in average vertical velocity, $V_y$, with depth) $dV_y/dy$, which is approximately the rate of porosity change, $d\Phi/dt$ (using eq 44 of Goren et al [2011]). Intense dilation and compaction rates (dark and light shading, respectively in the middle panel) frequently occur in adjacent layers. This is because when one row dilates by grains climbing on top of each other, the layer above it tends to compact due to the climbing grains jamming into it. Comparison of the top and middle panels show that zones of dilation and compaction occur at the same locality as the shear bands.

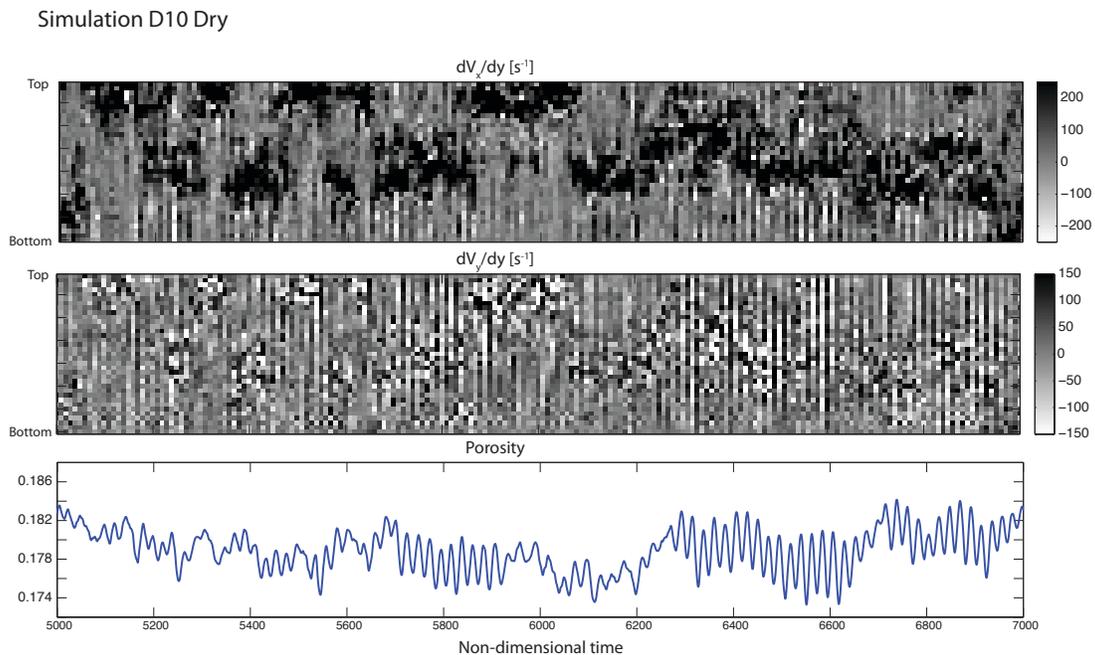

**Fig 2. Simulation D10 Dry. Upper panel: shading shows horizontally averaged shear strain, $dV_x/dy$, as function of depth in the granular layer (vertical axis) and non-dimensional time (horizontal axis). Black shading marks the location of shear bands. The shear bands do not persist in location, and instead move through the layer with time. Middle panel: rate of dilation (dark shading) and compaction (light shading), $dV_y/dy$, as function of depth and time. Both dilation (positive) and compaction (negative) correlate with rapid shear. Bottom panel: porosity as function of non-dimensional time. The duration recorded by the figure corresponds to a shear distance of ~240 grains at shear velocity $V_{sh}$.**

Figure 3 shows time-averaged profiles of normalized shear velocity vs. depth for several simulations. In simulation D10 Dry (dotted curve) the profile is nearly linear. The linearity indicates a time-averaged uniform shear strain rate throughout the system, even though most of the instantaneous profiles did show localization.

This confirms that shear bands in the dry system have no tendency to preferentially occur at any particular position in the system, nor to remain at any position for much strain.

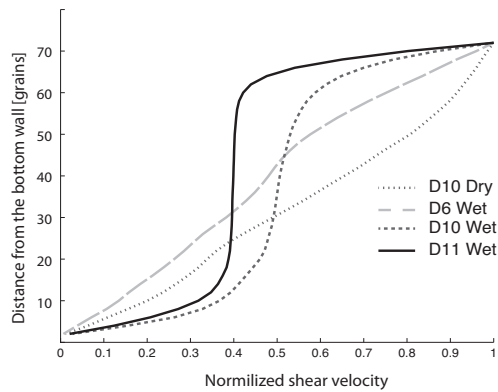

**Fig 3. Time-averaged shear velocity profiles for several simulations. A dry simulation (D10 Dry) and a high internal permeability wet simulation (D6 Wet) show linear profiles, while wet simulations with small internal permeability show persistent localization on the top and bottom boundaries. The vertical axis shows distance from the bottom wall in units of average grain diameter.**

**LOCALIZATION IN WET DRAINED GRANULAR LAYERS**

Figure 4 shows a simulation (D10 Wet) equivalent to that shown in Figure 2 but with fluid. In addition to the panels in Figure 2, Figure 4 includes a third panel that shows the pore pressure deviation from hydrostatic value, $P$ (dark shading for $P>0$ and light shading for $P<0$). Pressure deviations are created by dilation and compaction, and are diffused by fluid flow. As a consequence, pressure variations exhibit a much longer length scale than variations in $dV_y/dy$. Whereas the dry case (Figure 2) had 'wandering' shear bands, the wet simulation shows localization alternating between the top and bottom boundaries, with no significant shear bands forming in the interior of the layer. The short-dashed line in Figure 3 confirms that the time-averaged shear deformation is localized along the two boundaries of the system.

Localization along the boundaries in wet drained simulations results from the effect of the pore fluid pressure on the effective stress. A precondition for the formation of a shear band in granular matter is dilation [Marone 1991; Desrues et al., 1996; Makedonska et al., 2011]. Upon incipient dilation, the non-hydrostatic component of the pore pressure becomes negative and as a result the effective stress increases. This dilation hardening process resists further dilation and shear. Therefore, dilation and shear will preferentially take place in locations where the decrease in pore pressure does not significantly increase the effective stress, i.e. its magnitude is relatively small.

In our simulations, the boundaries of the layer are kept at a constant $P=0$ value by the drained boundary conditions. Goren et al. [2011] showed that, for the conditions that apply in these simulations, a balance between dilation rate and fluid flow governs non-hydrostatic pressures. As a consequence, dilation close to the boundaries will generate smaller magnitudes of negative pore pressure with respect to dilation further away from the boundaries. In order to quantify this effect, consider an internal position within the system, at a distance $y_b$ from a drained boundary. For a shear band to form at this position, there would be some initial dilation at a rate $d<\Phi>/dt$, where $<\Phi>$ is the average porosity within the layer. According to eq (51) from Goren et al [2011], the peak pore pressure difference, $\Delta P$, generated between this position and the drained boundary would be approximately

$$\Delta P = -\frac{\eta(\varsigma^2 - y^2)}{2k_0}\frac{d<\phi>}{dt} \sim -\frac{\eta\varsigma y_b}{2k_0}\frac{d<\phi>}{dt} \quad (1)$$

where $\eta$ is the fluid viscosity, $k_0$ is the mean permeability of the system, and $y = +\varsigma, -\varsigma$ are the positions of the drained boundaries – so that $y_b = \varsigma - y$, or $y_b = y - \varsigma$ and the last equality holds for positions close to the boundaries, where $y_b << \varsigma$. Equation (1) reveals that the magnitude of the pressure does not depend on applied normal stress, $\sigma_n$, but does depend on the dilation rate $d<\phi>/dt$, the distance to the boundary, $y_b$, and permeability, $k_0$.

We estimate that a shear band cannot form if the magnitude of the negative pressure perturbation generated during its formation (at the dilational stage) is greater than $-0.5\sigma_n$. Using equation (1), we can predict then that the largest distance from the boundary, $y_b$, along which a shear band can form, will be

$$y_b \leq \frac{k_0 \sigma_n}{\eta \varsigma d<\phi>/dt} \quad (2)$$

Our simulations show that localization on the boundaries occurs when the normal stress, $\sigma_n$, is small, the shear velocity, $V_{sh}$, is high (i.e. $\varsigma d<\phi>/dt$ is high), and the internal permeability, $k_0$, is small, as predicted by equation (2). For the conditions of simulations D10 Wet and D11 wet, and the value of dilation rate observed in Fig. 4, equation (2) predicts that shear bands should only form within about 3-8 grain diameters of the boundary, which is in good agreement with Figures 3 and 4.

Equation (2) and our simulations further show that when the normal stress is high, the shear velocity is small, or the internal permeability is large, shear bands can form at any distance from the boundaries because the magnitudes of the generated pore pressure deviations are always negligible with respect to the applied normal stress. Indeed, the long dashed curve in Figure 3 shows that simulation D6 Wet, that was conducted with the same parameters as D10 Wet, but with internal permeability greater by two orders of magnitude, have a close to uniform distribution of shear bands throughout the layer, as depicted by its close to linear averaged normalized shear velocity profile. The localization pattern in simulation D6 Wet is therefore more similar to a dry system.

Note that a positive pressure resulting from a rapid local compaction of a shear band will lead to a decrease in effective normal stress, which will reduce the resistance to shear along the shear band. Therefore positive $\Delta P$ from compaction will be largest for, and therefore favor, shear bands that form far from the boundary. The fact that shear band positions seem to be explained by equation (2) indicates that it is the dilational (hardening) phase which controls where shear bands can form.

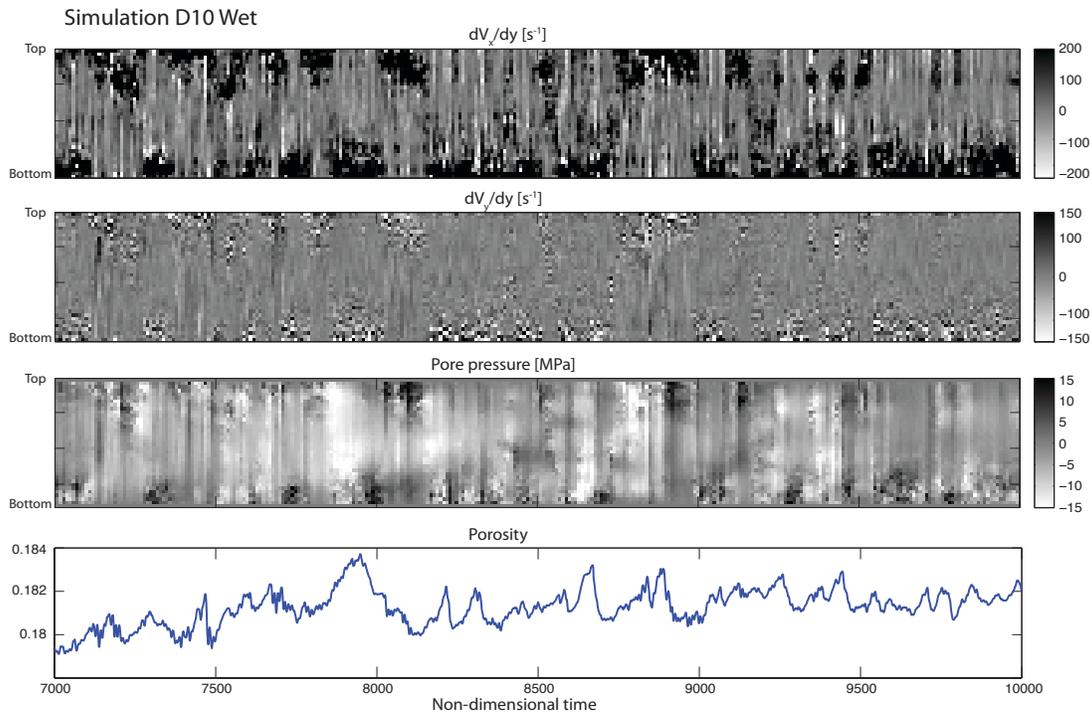

**Fig 4. Simulation D10 Wet. Top two panels: shear strain, $dV_x/dy$, and dilation rate, $dV_y/dy$, as described in Figure 2. Zones of alternating dilation (positive) and compaction (negative) correlate with shear bands. Third panel: Pore fluid pressure in excess of hydrostatic values. Dark shadings mark positive pressure and light shadings mark negative pressure. Bottom panel: porosity as function of non-dimensional time. The duration recorded by the figure corresponds to a shear distance of ~360 grains at shear velocity $V_{sh}$.**

## CONCLUSIONS

We conclude that pore fluid pressure has a significant effect on shear localization in low permeability granular systems as it leads to preferential shearing close to the drained boundaries. The reason for this behavior is that dilational hardening makes it preferable to dilate and shear in locations where the reduction of the pore pressure and the increase of the effective stress, during incipient dilation, can be relieved rapidly. Indeed, when the boundaries are more permeable than the interior, as is the case in many natural fault gouge layers, dilation close to the boundaries generates smaller magnitude of non-hydrostatic negative pore pressure with respect to dilation along an interior layer.

We further propose that such a permeability structure in natural fault gouge will favor shear localization close to the gouge boundaries. As a result of continued wear during shear, the gouge layer will expand at the expense of the bounding breccia by converting wall breccia into gouge material. A new breccia layer will be formed further away from the shear zone due to shear damage. This mechanism allows for the shear zone to migrate into the host rock through repeated earthquake events.